# Rank Energy Statistics in the Context of Change Point Detection

Amanda Ng

The Bronx High School of Science

Bodhisattva Sen

Mentor; Department of Statistics, Columbia University

**Introduction**

Change point analysis is the problem of detecting abrupt changes in data when a property of the time series changes. For a given observed time series, the instances where distributional changes occur are called change points. Energy distance is the statistical distance between two probability distributions. It can be used to test for equal distribution and the goodness-of-fit test and has numerous applications, including hierarchical clustering, testing multivariate normality, gene selection, and microarray data analysis.

In this paper, I will consider the rank energy statistic in the context of change point detection, rather than applying only energy distance to change point analysis. Using the rank energy statistic enables the detection of all departures from $H_0$, where the distributions are the same. This algorithm performs multiple change point analysis in a broad setting in which the observed distributions and number of change points are unspecified, rather than assume that the series follows a parametric model or there is only one change point, as many works in this area of study assume. This proposed algorithm does not require previous knowledge nor supplementary analysis, unlike in [3], [4], and [5].

**Methods**

Here I will consider the rank energy statistic, a distribution-free goodness-of-fit measure using E-statistics, in the context of estimating the location of a change point. The rank energy statistic is used for testing the equality of two multivariate distributions.

I describe my method below, where $\mu_m^X$ and $\mu_n^Y$ denote the empirical distributions on $D_m^X := \{X_1,..., X_m\}$ and $D_n^Y := \{Y_1, ..., Y_n\}$ respectively. Let $\mu_{m,n}^{X,Y} := (m+n)^{-1}(m\mu_n^X + n\mu_n^Y)$ and let $H_{m+n}^d := \{h_1^d, ..., h_{m+n}^d\} \subset [0,1]^d$ denote the fixed sample multivariate ranks.

The empirical distribution on $H_{m+n}^d$ weakly converges to $U^d$ as $\min(m,n) \to \infty$, where $H_{m+n}^d$ is the d-dimensional Halton sequence for $d \geq 2$ and $\{i/(m+n) : 1 \leq i \leq m+n\}$ for $d=1$.

I will use $\widehat{R}_{m,n}^{X,Y}(\cdot)$ to denote the joint empirical rank map corresponding to the transportation of $\mu_{m,n}^{X,Y}$ to the empirical distribution $H_{m+n}^d$. The change point detection using a rank energy statistic is defined as:

$$\widehat{\mathcal{E}}(X_n, Y_m; \alpha) = \frac{2}{mn} \sum_{i=1}^{n} \sum_{j=1}^{m} |\widehat{R}_{m,n}^{X,Y}(X_i) - \widehat{R}_{m,n}^{X,Y}(Y_j)|^\alpha - \binom{n}{2}^{-1} \sum_{1 \leq i \leq k \leq n} |\widehat{R}_{m,n}^{X,Y}(X_i) - \widehat{R}_{m,n}^{X,Y}(X_k)|^\alpha$$

$$- \binom{m}{2}^{-1} \sum_{1 \leq j \leq k \leq m} |\widehat{R}_{m,n}^{X,Y}(Y_j) - \widehat{R}_{m,n}^{X,Y}(Y_k)|^\alpha \quad (1)$$

$$\widehat{Q}(X_n, Y_m; \alpha) = \frac{mn}{m+n} \widehat{\mathcal{E}}(X_n, Y_m; \alpha) \quad (2)$$

$$(3)$$

$$(\hat{\tau}, \hat{\kappa}) = \underset{(\tau,\kappa)}{argmax}\ \hat{Q}(X_\tau, Y_\tau(\kappa); \alpha)$$

Observe that the right hand side of (1) can be viewed as the rank energy statistic; this measure is based on Euclidean distances between sample elements. This rank energy statistic can also be viewed as a rank-transformed version of the empirical energy measure in [1]. (2) denotes the scaled sample empirical divergence measure that leads to a consistent approach for estimating change point locations. Let $Z_1, ..., Z_T \in \mathbb{R}^d$ be an independent sequence of observations and let $1 \leq \tau < \kappa \leq T$ be constants. Let the following sets $X_\tau = \{Z_1, Z_2, ..., Z_\tau\}$ and $Y_\tau(\kappa) = \{Z_{\tau+1}, Z_{\tau+2}, ..., Z_\kappa\}$. A change point location $\hat{\tau}$ is then estimated in (3). This procedure can also be seen as a ranked version of the change point detection algorithm used in [2].

All the methods described above have been implemented using the R software. The relevant codes are available in GitHub under the repository found at "**Amanda-Ng/recp**".

**Financial Data**

Here I will apply the proposed algorithm to the monthly log returns of Cisco Systems Inc., a multinational technology conglomerate in the development and manufacturing of networks, from April 1990 to January 2012. I estimated two significant change points, each with estimated p-values below 0.03. The series shown in the figure below indicates the approximate change points at the vertical green lines, which are located at April 2000 and October 2002.

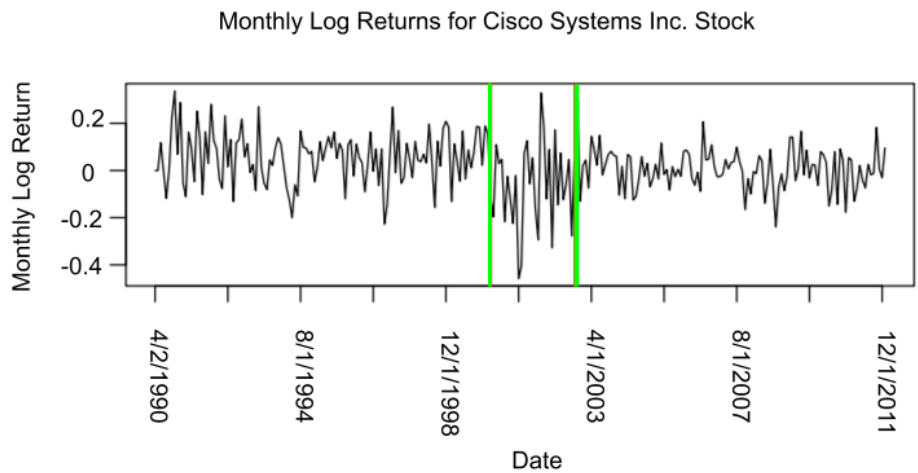

Figure 1: Monthly log returns for Cisco Systems Inc. stock from April 1990 to January 2012; the proposed algorithm estimates 2 significant distributional changes at the green vertical lines denoting April 2000 and October 2002.

The first change point at April 2000 aligns with the company's acquisition of Pirelli Optical Systems to counter competitors Nortel and Lucent. This acquisition enabled Cisco to provide its customers with a more complete network infrastructure and lower network costs. The second significant change point at October 2002 corresponds to the end of a period of highly aggressive pursuits in engaging markets. During this period, Cisco developed a multi-billion dollar network for Shanghai, which later became China's largest urban communications system.